\documentclass[twocolumn,letterpaper,prl,showpacs,superscriptaddress]{revtex4}
\usepackage{graphicx}
\usepackage{bm}
\usepackage{amsmath}
\usepackage{amssymb}
\newlength{\twocolumnwidth}

\newlength{\figwidth} 
\newcommand{\IDWickCaus}{1104.3764}
\newcommand{\IDQDynResp}{1104.3809}
\newcommand{\IDPFunc}{1104.3825}

\newcommand{\PFunc}{arXive:\IDPFunc\ (2011)}

\begin{document} 
\title{%
Operator ordering and causality} 
\author{L.\ I.\ Plimak} 
\affiliation{Institut f\"ur Quantenphysik, Universit\"at Ulm, 
D-89069 Ulm, Germany.} 
\author{S.\ Stenholm} 
\affiliation{Institut f\"ur Quantenphysik, Universit\"at Ulm, 
D-89069 Ulm, Germany.} 
\affiliation{Physics Department, Royal Institute of Technology, KTH, Stockholm, Sweden.} 
\affiliation{Laboratory of Computational Engineering, HUT, Espoo, Finland.} 
\date{\today} 
\begin{abstract} 
It is shown that causality violations [M.\ de Haan, Physica {\bf 132}A, 375, 397 (1985)], emerging when the conventional definition of the time-normal operator ordering [P.L.Kelley and W.H.Kleiner, Phys.Rev.\ {\bf 136}, A316 (1964)] is taken outside the rotating wave approximation, disappear when the amended definition [L.P.\ and S.S., Annals of Physics, {\bf 323}, 1989 (2008)] of this ordering is used. 
\end{abstract}
\pacs{XXZ}
\maketitle 
{\em Introduction.---\/} Causality in photocounting is a longstanding issue \cite{FermShir,deHaan,BykTat,Tat,MilonniEtc}. No one really doubts that field radiation and detection is a causal process (cf, e.g., \mbox{Refs.\ \cite{MilonniEtc}}), but to identify universal causal quantities measured by a macroscopic detector of arbitrary design remains an open question. Such quantities naturally emerge in {\em response formulation\/} of quantum electrodynamics (QED) \cite{API,APII,APIII,Theory,PFunc}. In this letter we consider causality properties of these quantities and show that their use eliminates all causality issues from the theory. 

Referring the reader for details to the cited papers, here we only outline the key points. In Glauber's photodetection theory \cite{GlauberPhDet,KelleyKleiner,MandelWolf}, spectral properties of the detected field are determined by the quantum average $\ensuremath{\big\langle 
{\hat{\mathcal E}}^{(-)}(x,t)
{\hat{\mathcal E}}^{(+)}(x,t')
\big\rangle} $, where ${\hat{\mathcal E}}^{(\pm)}(x,t)$ denote frequency-positive or negative parts of the {Heisenberg}\ field operator ${\hat{\mathcal E}}(x,t)={\hat{\mathcal E}}^{(+)}(x,t)
+{\hat{\mathcal E}}^{(-)}(x,t)$, and $x$ comprises all field arguments except time. De Haan \cite{deHaan} and later Bykov and Tatarskii \cite{BykTat} remarked that $\ensuremath{\big\langle 
{\hat{\mathcal E}}^{(-)}(x,t)
{\hat{\mathcal E}}^{(+)}(x',t')
\big\rangle} $ is not a causal quantity, and suggested that the actual measured quantity should be the time-normal (TN) average 
{\begin{align}{{
 \begin{aligned} 
\ensuremath{\big\langle{\mathcal T}{\mbox{\rm\boldmath$:$}} 
{\hat{\mathcal E}}(x,t){\hat{\mathcal E}}(x',t')
{\mbox{\rm\boldmath$:$}}
\big\rangle} . 
\end{aligned}}}%
\label{eq:34BJ} 
\end{align}}%
The symbol ${\mathcal T}{\mbox{\rm\boldmath$:$}}\cdots{\mbox{\rm\boldmath$:$}}$ denotes the TN operator ordering of Kelley and Kleiner (KK) \cite{KelleyKleiner,MandelWolf}. Quantity (\ref{eq:34BJ}) differs from $2\mathrm{Re}\ensuremath{\big\langle 
{\hat{\mathcal E}}^{(-)}(x,t)
{\hat{\mathcal E}}^{(+)}(x',t')
\big\rangle} $ by nonresonant terms. However, Tatarskii \cite{Tat} pointed out that quantity (\ref{eq:34BJ}) may also exhibit acausal behaviour. Here we show that the idea of de Haan, Bykov and Tatarskii is basically correct. The problem is with the KK definition of the TN ordering which needs generalisation beyond the rotating wave approximation (RWA). By using the amended definition of \mbox{Ref.\ \cite{APII}} all causality problems are eliminated. 

{\em Characteristic properties of the TN ordering.---\/} 
If quantities measured by a detector are universal, i.e., independent of the detector, they should be identifiable in a QED theory of a field source. 
Let 
{\begin{align}{{
 \begin{aligned} 
& \hat E(x,t) = \sum_{\kappa}\sqrt{\frac{\hbar}{2\omega _{\kappa }}}\,
u_{\kappa}(x)\hspace{1pt}\text{e}^{-i\omega _{\kappa} t}\hat a _{\kappa} + 
\textrm{H.c.}
\end{aligned}}}%
\label{eq:27BA} 
\end{align}}%
be a Hermitian field operator, described by the standard bosonic creation and annihilation\ pairs $\hat a_{\kappa}^{\dag},\hat a_{\kappa}$, 
with $u_{\kappa}(x)$ being complex mode functions. The field interacts with a quantum source according to the Hamitonian, 
{\begin{gather}{{
 \begin{gathered} 
\hat H_{\textrm{int}}(t) = 
- \int dx \hat J(x,t)
\hat E(x,t) 
, 
\end{gathered}}}%
\label{eq:26AZ} 
\end{gather}}%
The nature of the mode index $\kappa $, of the free current operator $\hat J(x,t)$ and of the source Hamiltonian (occuring implicitly) may be arbitrary. 
By definition, $\hat E(x,t)$ and $\hat J(x,t)$ are the interaction-picture (free) operators; the same operators in the {Heisenberg}\ picture are ${\hat{\mathcal E}}(x,t)$ and ${\hat{\mathcal J}}(x,t)$. Free-field operators and the interaction Hamiltonian is all one needs to construct the standard nonstationary perturbation theory \cite{Schweber}. This makes the {Heisenberg}\ operators formally defined, at least in perturbative terms. 

We define the (generalised) TN operator ordering postulating that, 1) {\em under the TN ordering, classical radiation laws apply directly to {Heisenberg}\ operators\/}. Formally, this is expressed by the relation between the TN averages of the field and current operators, 
{\begin{multline}\hspace{0.4\columnwidth}\hspace{-0.4\twocolumnwidth} 
\ensuremath{\big\langle{\mathcal T}{\mbox{\rm\boldmath$:$}} 
{\hat{\mathcal E}}(x_1,t_1)\cdots{\hat{\mathcal E}}(x_m,t_m) 
{\mbox{\rm\boldmath$:$}}\big\rangle} = \int dx'_1 dt'_1\cdots dx'_m dt'_m 
\\ \times 
\Delta _{\text{R}}(x_1,x_1',t_1-t'_1)\cdots
\Delta _{\text{R}}(x_m,x_m',t_m-t'_m)
\\ \times 
\ensuremath{\big\langle{\mathcal T}{\mbox{\rm\boldmath$:$}} 
{\hat{\mathcal J}}(x_1',t_1')\cdots{\hat{\mathcal J}}(x_1',t_m') 
{\mbox{\rm\boldmath$:$}}\big\rangle} , 
\hspace{0.4\columnwidth}\hspace{-0.4\twocolumnwidth} 
\label{eq:25AY} 
\end{multline}}%
where $ \Delta _{\text{R}}(x,x',t-t')$ is Kubo's linear response function of the free field, 
{\begin{multline}\hspace{0.4\columnwidth}\hspace{-0.4\twocolumnwidth} 
\Delta _{\text{R}}(x,x',t-t') = \frac{i}{\hbar }\theta(t-t')\ensuremath{\big[
\hat E(x,t),\hat E(x',t')
\big]} \\ 
= -2\theta(t-t')\mathrm{Im}\sum_{\kappa} u_{\kappa}(x)u_{\kappa}^*(x')\text{e}^{-i\omega _{\kappa} (t-t')}
. 
\hspace{0.4\columnwidth}\hspace{-0.4\twocolumnwidth} 
\label{eq:30BD} 
\end{multline}}%
The averaging in (\ref{eq:25AY}) is over the {Heisenberg}\ (initial) state of the system. The latter is assumed to factorise into the vacuum state of all oscillators (denoted $\left|0\right\rangle$) and an arbitrary state of the quantum source. 
We retain the conventional notation for the TN ordering, implying that the KK definition is a resonance approximation to it (which is indeed the case, see below). 

Linear response functions of free bosonic fields and of the corresponding classical ones are identical \cite{API}. Hence eq.\ (\ref{eq:25AY}) exactly emulates a classical formula for stochastic averages of a field radiated by a random current into empty space (vacuum). Such formula is found from eq.\ (\ref{eq:25AY}) by dropping hats and replacing TN averages by classical stochastic averages. 
Note that condition 1 warrants explicit relativistic causality in radiation and propagation of the field but says nothing about causality properties of the TN ordering itself. 

Furthermore, 2) {\em for free electromagnetic operators, the TN ordering coincides with the conventional normal ordering\/} \cite{MandelWolf}. 
This is a consistency requirement: with ${\hat{\mathcal J}}(x,t)=0$ eq.\ (\ref{eq:25AY}) turns into, 
{\begin{align}{{
 \begin{aligned} 
\ensuremath{\big\langle 0\big|
{\mathcal T}{\mbox{\rm\boldmath$:$}} 
\hat E(x_1,t_1)\cdots\hat E(x_m,t_m) 
{\mbox{\rm\boldmath$:$}}
\big|0\big\rangle} = 0 . 
\end{aligned}}}%
\label{eq:76JM} 
\end{align}}%
This relation is enforced by condition 2. 

{\em Time-normal ordering beyond the RWA.---\/} 
Characteristic conditions 1 and 2 are supplemented by formal ones:
3) {\em the operation ${\mathcal T}{\mbox{\rm\boldmath$:$}}\cdots{\mbox{\rm\boldmath$:$}}$ is polylinear\/}, 4) {\em any commuting factor may be taken out of the ${\mathcal T}{\mbox{\rm\boldmath$:$}}\cdots{\mbox{\rm\boldmath$:$}}$ symbol\/}, and 5) ${\mathcal T}{\mbox{\rm\boldmath$:$}}\hat\openone{\mbox{\rm\boldmath$:$}}=\hat\openone$. The TN product of $m$ arbitrary operators ${\hat{\mathcal A}}_k(t),\ k=1,\cdots,m$, obeying conditions 1--5 \footnote{While conditions 1--5 appear to specify (\ref{eq:32CT}) uniquely (cf. \mbox{Ref.\ \cite{API}}, appendix 2), we do not claim this as a theorem.}, is defined as \cite{APII,APIII}, 
{\begin{multline}\hspace{0.4\columnwidth}\hspace{-0.4\twocolumnwidth} 
{\mathcal T}{\mbox{\rm\boldmath$:$}}{\hat{\mathcal A}}_1(t_1)\cdots{\hat{\mathcal A}}_m(t_m){\mbox{\rm\boldmath$:$}} \\ 
= \int dt_1'\cdots dt_m'
T_C \prod_{k=1}^m
\ensuremath{\big[
\delta ^{(+)}(t_k-t'_k){\hat{\mathcal A}}_{k+}(t'_k) 
\\ 
+ 
\delta ^{(-)}(t_k-t'_k){\hat{\mathcal A}}_{k-}(t'_k)
\big]} . 
\hspace{0.4\columnwidth}\hspace{-0.4\twocolumnwidth} 
\label{eq:32CT} 
\end{multline}}%
The kernels $\delta ^{(\pm)}$
are the frequency-positive and negative\ parts of the delta-function, (cf.\ \mbox{Ref.\ \cite{APII}}, appendix 2)
{\begin{align}{{
 \begin{aligned} 
\delta ^{(\pm)}(t) 
= \ensuremath{\pm\frac{1}{2\pi i(t\mp i0^+)}} , 
\end{aligned}}}%
\label{eq:40KW} 
\end{align}}%
and the $T_C$, or closed-time-loop, ordering \cite{SchwingerC} is a way of writing the {\em double-time-ordered\/} operator structure (known, e.g., from the photodetection theory 
\cite{KelleyKleiner,MandelWolf}), 
{\begin{multline}\hspace{0.4\columnwidth}\hspace{-0.4\twocolumnwidth} 
\bar T
{\hat{\mathcal A}}_{1}(t_1)\cdots{\hat{\mathcal A}}_{m}(t_m)
\,T 
{\hat{\mathcal B}}_{1}(t'_1)\cdots{\hat{\mathcal B}}_{n}(t'_n) 
\\ \equiv 
T_C 
{\hat{\mathcal A}}_{1-}(t_1)\cdots{\hat{\mathcal A}}_{m-}(t_m)
{\hat{\mathcal B}}_{1+}(t'_1)\cdots{\hat{\mathcal B}}_{n+}(t'_n) 
, 
\hspace{0.4\columnwidth}\hspace{-0.4\twocolumnwidth} 
\label{eq:30CR} 
\end{multline}}%
where $T$ is the standard time ordering of operators and $\bar T$ is the ``reverse'' ordering. The $_{\pm}$ indices serve only for ordering purposes and otherwise should be disregarded. 

Polylinearity of (\ref{eq:32CT}) (property 3) is inherited from the $T_C$-ordering. Property 4 is a consequence of the relation, 
{\begin{align}{{
 \begin{aligned} 
\delta ^{(+)}(t)
+\delta ^{(-)}(t) = \delta (t). 
\end{aligned}}}%
\label{eq:24CK} 
\end{align}}%
Property 5 is a specification of (\ref{eq:32CT}) for $m=0$. Property 2 is verified in \mbox{Ref.\ \cite{API}}. The one most difficult to demonstrate is property 1. Its formal proof \cite{Theory,PFunc} employs heavy-duty machinery of quantum field theory. However, the basic idea is fairly simple. The starting point of the proof is the {\em wave quantisation formula\/} \cite{Corresp}, 
{\begin{multline}\hspace{0.4\columnwidth}\hspace{-0.4\twocolumnwidth} 
\ensuremath{\big[
\hat E(x,t),\hat E(x',t')
\big]} = -i\hbar \ensuremath{\big[
\Delta _{\text{R}}(x,x',t-t') \\ 
-\Delta _{\text{R}}(x',x,t'-t)
\big]} . 
\hspace{0.4\columnwidth}\hspace{-0.4\twocolumnwidth} 
\label{eq:20AT} 
\end{multline}}%
It is found inverting Kubo's formula (\ref{eq:30BD}). Equation (\ref{eq:20AT}) induces restructuring firstly of Wick's theorem, and consequently of the whole standard perturbative approach of the quantum field theory. Equation (\ref{eq:25AY}) is a rigorous form of eq.\ (74) in paper \cite{PFunc}. 

To extend (\ref{eq:32CT}) to fermionic operators one may assume that they always occur multiplied by generators of an auxiliary Grassmann algebra \cite{APIII,VasF,Beresin}, 
{\begin{align}{{
 \begin{aligned} 
\hat A_k(t_k) = \gamma_k \hat F_k(t_k) . 
\end{aligned}}}%
\label{eq:29CQ} 
\end{align}}%
Such combinations behave under orderings as bosonic operators \cite{APIII}. Quantum fields are included by making all field ``labels'' explicit. For instance, in spinor electrodynamics, $\hat A_k(t_k)=\hat A_{\mu _k}({\mbox{\rm\boldmath$r$}}_k,t_k),\gamma _k\hat \psi _{\sigma _k}({\mbox{\rm\boldmath$r$}}_k,t_k)$, where $\hat A_{\mu }({\mbox{\rm\boldmath$r$}},t)$ is the 4-vector electromagnetic potential and $\hat \psi _{\sigma }({\mbox{\rm\boldmath$r$}},t)$ is the electron-positron field. Other quantum fields may be included similarly. 
Equation (\ref{eq:32CT}) thus extends the KK definition in two ways: beyond the RWA and to arbitrary quantised fields including fermions. 

{\em Time-normal ordering under the RWA.---\/} 
Separation of the frequency-positive and negative\ parts of a function, occuring in Kelley-Kleiner's definition of the TN ordering, may be written as an integral transformation, (see, e.g., \mbox{Ref.\ \cite{APII}}, appendix 2) 
{\begin{align}{{
 \begin{aligned} 
f ^{(\pm)}(t) 
= \int_{-\infty}^{\infty} dt' \delta ^{(\pm)}(t-t') f (t'). 
\end{aligned}}}%
\label{eq:39KV} 
\end{align}}%
In (\ref{eq:32CT}), the $T_C$-ordering applies to entire operators, and not to their frequency-positive and negative\ parts (i.e., $T_C$ first, $^{(\pm)}$ second). The KK definition emerges by changing the order of operations ($^{(\pm)}$ first, $T_C$ second) \cite{APII}. This results in a {\em resonance approximation\/} to (\ref{eq:32CT}): 
{\begin{multline}\hspace{0.4\columnwidth}\hspace{-0.4\twocolumnwidth} 
{\mathcal T}{\mbox{\rm\boldmath$:$}}{\hat{\mathcal A}}_1(t_1)\cdots{\hat{\mathcal A}}_m(t_m){\mbox{\rm\boldmath$:$}} \\ 
\approx 
T_C \prod_{k=1}^m
\ensuremath{\big[
{\hat{\mathcal A}}_{k+}^{(+)}(t_k) 
+ 
{\hat{\mathcal A}}_{k-}^{(-)}(t_k)
\big]} , 
\hspace{0.4\columnwidth}\hspace{-0.4\twocolumnwidth} 
\label{eq:35BK} 
\end{multline}}%
which indeed coincides with the KK definition. 

\begin{figure} [t]
\begin{center} 
\includegraphics[width=0.85\figwidth]
{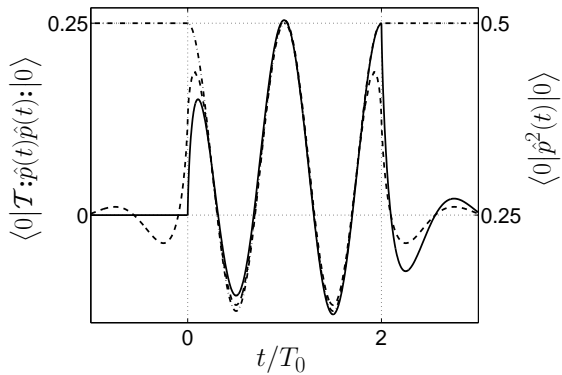}
\end{center} 
\caption{The time-normal average 
$\big\langle 0\big|\mathcal{T}{\mbox{\rm\boldmath$:$}}
\hat p(t)\hat p(t)
{\mbox{\rm\boldmath$:$}}\big| 0\big\rangle$ (solid line), the KK approximation to it (dashed line) and the average $\ensuremath{\langle 0| 
\hat p^2(t)
|0\rangle} $ (dash-dotted line) in units of $\hbar m\omega _0$. 
The graph of $\ensuremath{\langle 0| 
\hat p^2(t)
|0\rangle} $ is shifted vertically by $\hbar m\omega _0/4$.%
}
\label{fig:XX} 
\end{figure}

{\em Parametric oscillator.---\/} 
The simplest example of a system which exhibits a causality violation with definition (\ref{eq:35BK}) is the parametric oscillator, 
{\begin{gather}{{
 \begin{gathered} 
\hat H(t) = \frac{\hat p^2(t)}{2m} + \frac{m\omega^2(t)\hat x^2(t)}{2} . 
\end{gathered}}}%
\label{eq:26CM} 
\end{gather}}%
Its frequency $\omega _0$ is reduced by half at $t = 0$ and restored to its initial value at $t=2T_0 = 4\pi /\omega _0$: 
{\begin{align}{{
 \begin{aligned} 
\omega (t) = 
\left\{\begin{array}{cl}
\displaystyle\frac{\omega _0}{2} , & 0<t<2T_0, \\
\omega _0, & t<0\mathrm{\ or\ }t>2T_0. 
\end{array}\right.
\end{aligned}}}%
\label{eq:27CN} 
\end{align}}%
The initial ({Heisenberg}) state of the oscillator is vacuum (defined with respect to $\omega _0$). For $\hat p(t)$ we have, 
{\begin{align}{{
 \begin{aligned} 
\hat p(t) = 
\begin{cases}\displaystyle 
\hat p \cos\frac{\omega _0t}{2} - \frac{m\omega _0\hat x}{2} \sin \frac{\omega _0t}{2}, 
& 0<t<2T_0, \\ \displaystyle 
\hat p \cos{\omega _0t} - m\omega _0\hat x \sin {\omega _0t}, 
& t<0\mathrm{\ or\ }t>2T_0. 
\end{cases}
\end{aligned}}}%
\label{eq:d52a} 
\end{align}}%
The quantity $\ensuremath{\big\langle 0\big|
\mathcal{T}{\mbox{\rm\boldmath$:$}}
\hat p(t)\hat p(t')
{\mbox{\rm\boldmath$:$}}
\big|0\big\rangle} 
$ for $t=t'$ calculated using eqs.\ (\ref{eq:32CT}) and (\ref{eq:35BK}) is drawn in Fig.\ \ref{fig:XX} with solid and dashed lines, respectively. The former is zero for $t<0$, whereas the latter is nonzero for all times. Using eq.\ (\ref{eq:35BK}) indeed resulted in a causality violation. 

Used here solely as an illustration, the parametric oscillator appears to be of much interest by itself. The physical motivation for calculating $\ensuremath{\big\langle 0\big|
\mathcal{T}{\mbox{\rm\boldmath$:$}}
\hat p(t)\hat p(t')
{\mbox{\rm\boldmath$:$}}
\big|0\big\rangle} 
$ is that it describes a {\em continuous measurement of fluctuations of the quantised momentum by means of electromagnetic interaction\/}. The quantum current associated with the particle is ${\hat{\mathcal J}}(t)=e\hat p(t)/m$, where $e$ is its charge. Radiation of this current is observed by a remote detector. In essense, Hamiltonian (\ref{eq:26CM}) is postulated as an effective one for the {Heisenberg}\ current operator ${\hat{\mathcal J}}(t)=e\hat p(t)/m$. Figure \ref{fig:XX} demonstrates two things. Firstly, the time-normal ordering matters. For comparison, we also draw in Fig.\ \ref{fig:XX} the quantity $\ensuremath{\langle 0| 
\hat p^2(t)
|0\rangle} $ (dash-dotted line) \footnote{Vacuum $\left|0\right\rangle$ is a squeezed state of the oscillator with frequency $\omega _0/2$. Oscillations of $\ensuremath{\langle 0| 
\hat p^2(t)
|0\rangle} $ reflect its evolution. See, e.g., W.P.Schleich, {\em Quantum Optics in Phase Space\/} (Wiley, 2001), p.\ 119.}. Clearly the exact and the KK results for $\ensuremath{\big\langle 0\big|
\mathcal{T}{\mbox{\rm\boldmath$:$}}
\hat p(t)\hat p(t')
{\mbox{\rm\boldmath$:$}}
\big|0\big\rangle} 
$ are much closer to each other than to $\ensuremath{\langle 0| 
\hat p^2(t)
|0\rangle} $. Secondly, all differences are limited to short transients at $t=0$ and $t=2T_0$. Up to a vertical shift, the beyond-the-RWA $\ensuremath{\big\langle 0\big|
\mathcal{T}{\mbox{\rm\boldmath$:$}}
\hat p(t)\hat p(t')
{\mbox{\rm\boldmath$:$}}
\big|0\big\rangle} 
$ and $\ensuremath{\langle 0| 
\hat p^2(t)
|0\rangle} $ quickly settle into the same pattern; for $T_0/2<t<2T_0$ they are indistinguishable. 
Furthermore, the reader may be surprised --- and even disturbed --- by the transient seen in Fig.\ \ref{fig:XX} for times beyond $2T_0$. It should not be overlooked that the oscillator returns to its vacuum state at $t=2T_0$ \footnote{Note also that $\ensuremath{\langle 0| 
\hat p^2(t)
|0\rangle}$ dutifully returns to its vacuum value $\hbar m\omega _0/2$ for $t\geq 2T_0$}. As suggested by A.Kaplan \cite{Kaplan}, this transient may be a toy case of the Unruh effect \cite{Unruh}. Indeed, since $\ensuremath{\langle 0|
{\hat{\mathcal J}}(t)
|0\rangle} =0$, radiation of the current ${\hat{\mathcal J}}(t)$ is spontaneous emission. The transient reflects its non-instantaneity, which in turn is due to the time-energy uncertainty relation. 
However, to claim results in Fig.\ \ref{fig:XX} as physical, one has to account for the radiation friction {\em and its fluctuations\/} disregarded in the effective Hamiltonian. 
We return to this discussion elsewhere. 

{\em ``No-peep-into-the-future'' theorem.---\/} 
We now show that {\em a TN product depends on the operators it comprises only for times not later than its latest time argument\/}. The question of causality of TN products thus reduces to that for quantum equations of motion.

Proof of this theorem is a simplified version of the proof of causality of quantum response functions in \mbox{Ref.\ \cite{APII}}. 
We break the integration in (\ref{eq:32CT}) into $m$ domains, labelled by $k = 1, \cdots, m$, such that, 
{\begin{align}{{
 \begin{aligned} 
t'_1,\cdots,t'_{k-1},t'_{k+1},\cdots,t'_n<t'_k . 
\end{aligned}}}%
\label{eq:22CH} 
\end{align}}%
Within the $k$th domain, the $k$th factor in the product in (\ref{eq:32CT}) is transformed as follows, (cf.\ eq.\ (\ref{eq:24CK}))
{\begin{multline}\hspace{0.4\columnwidth}\hspace{-0.4\twocolumnwidth} 
\int_{-\infty}^{\infty} dt'_k\ensuremath{\big[
\delta ^{(-)}(t_k-t'_k){\hat{\mathcal A}}_-(t'_k)
+\delta ^{(+)}(t_k-t'_k){\hat{\mathcal A}}_+(t'_k)
\big]} 
\\ 
= 
\int_{-\infty}^{\infty} dt'_k{\hat{\mathcal A}}_+(t'_k)\ensuremath{\big[
\delta ^{(-)}(t_k-t'_k)
+\delta ^{(+)}(t_k-t'_k)
\big]} 
\\ 
= {\hat{\mathcal A}}_+(t_k). 
\hspace{0.4\columnwidth}\hspace{-0.4\twocolumnwidth} 
\label{eq:23CJ} 
\end{multline}}%
Indeed, the latest operator in eq.\ (\ref{eq:30CR}) is positioned between the $\bar T$ and $T$-ordered products {\em irrespective of its C-contour index\/} (cf.\ eq.\ (110) in \mbox{Ref.\ \cite{APII}}). The integration over remaining variables is restricted to, 
{\begin{align}{{
 \begin{aligned} 
t'_1,\cdots,t'_{k-1},t'_{k+1},\cdots,t'_n<t_k, 
\end{aligned}}}%
\label{eq:25CL} 
\end{align}}%
making the ``no-peep-into-the-future'' theorem\ evident. 

\begin{figure}[t]
\begin{center} 
\includegraphics[width=0.65\figwidth]{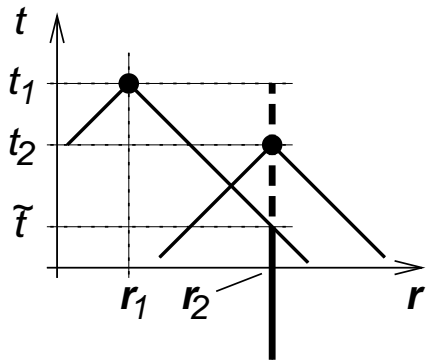}
\end{center} 
\caption{Space-time geometry of the integral in eq.\ (\ref{eq:37CY}). Points ${\mbox{\rm\boldmath$r$}}_1,t_1$ and ${\mbox{\rm\boldmath$r$}}_2,t_2$ are shown as bold dots, and their past light cones as slanted lines. The thick vertical line represents the ``integration path'' ${\mbox{\rm\boldmath$r$}}_2,t_2'$, $-\infty<t_2'<t_1$. It consists of two sections, positioned inside ($-\infty<t_2'<\tilde t$, solid) and outside ($\tilde t<t_2'<t_1$, dashed) of the past light cone of ${\mbox{\rm\boldmath$r$}}_1,t_1$. Other lines guide the eye.} 
\label{fig:RelCaus}
\end{figure}

{\em Relativistic causality.---\/} As a comparatively simple example, consider the pair time-normal product, 
{\begin{multline}\hspace{0.4\columnwidth}\hspace{-0.4\twocolumnwidth} 
{\mathcal T}{\mbox{\rm\boldmath$:$}}{\hat{\mathcal A}}_1({\mbox{\rm\boldmath$r$}}_1,t_1){\hat{\mathcal A}}_2({\mbox{\rm\boldmath$r$}}_2,t_2){\mbox{\rm\boldmath$:$}} 
= \int_{-\infty}^{\infty} dt'_1\int_{-\infty}^{\infty}dt'_2 T_C
\\ \times 
\ensuremath{\big[
\delta ^{(+)}(t_1-t'_1){\hat{\mathcal A}}_{1+}({\mbox{\rm\boldmath$r$}}_1,t'_1) 
+ 
\delta ^{(-)}(t_1-t'_1){\hat{\mathcal A}}_{1-}({\mbox{\rm\boldmath$r$}}_1,t'_1)
\big]} 
\\ \times 
\ensuremath{\big[
\delta ^{(+)}(t_2-t'_2){\hat{\mathcal A}}_{2+}({\mbox{\rm\boldmath$r$}}_2,t'_2) 
+ 
\delta ^{(-)}(t_2-t'_2){\hat{\mathcal A}}_{2-}({\mbox{\rm\boldmath$r$}}_2,t'_2)
\big]} . 
\hspace{0.4\columnwidth}\hspace{-0.4\twocolumnwidth} 
\label{eq:37CY} 
\end{multline}}%
Hereinafter we assume that $t_1>t_2$. In the spirit of the above proof, we break the integration into the two domains, $t'_1>t'_2$ and $t'_2>t'_1$. In both domains, time ordering is performed explicitly. Rearranging the terms and using (\ref{eq:24CK}) we find, 
{\begin{multline}\hspace{0.4\columnwidth}\hspace{-0.4\twocolumnwidth} 
{\mathcal T}{\mbox{\rm\boldmath$:$}}{\hat{\mathcal A}}_1({\mbox{\rm\boldmath$r$}}_1,t_1){\hat{\mathcal A}}_2({\mbox{\rm\boldmath$r$}}_2,t_2){\mbox{\rm\boldmath$:$}} 
\\ = \int_{-\infty}^{t_1} dt'_2
\ensuremath{\big[
\delta ^{(+)}(t_2-t'_2){\hat{\mathcal A}}_{1}({\mbox{\rm\boldmath$r$}}_1,t_1){\hat{\mathcal A}}_{2}({\mbox{\rm\boldmath$r$}}_2,t'_2) 
\\ + 
\delta ^{(-)}(t_2-t'_2){\hat{\mathcal A}}_{2}({\mbox{\rm\boldmath$r$}}_2,t'_2){\hat{\mathcal A}}_{1}({\mbox{\rm\boldmath$r$}}_1,t_1)
\big]} 
+ \ensuremath{\{
1\leftrightarrow 2
\}} . 
\hspace{0.4\columnwidth}\hspace{-0.4\twocolumnwidth} 
\label{eq:39DA} 
\end{multline}}%
The integral here is illustrated in Fig.\ \ref{fig:RelCaus}, where the space-time points ${\mbox{\rm\boldmath$r$}}_1,t_1$ and ${\mbox{\rm\boldmath$r$}}_2,t_2$ are drawn as bold dots, and the domain \mbox{${\mbox{\rm\boldmath$r$}}_1,t'_1,-\infty<t'_1<t_2$} (``integration path'') as a thick vertical line with solid and dashed sections. 

Now, what kind of ``no-peep-into-the-future'' theorem\ would one expect in relativity? Assuming that the dependence of operators on various perturbations is relativistically causal, it is sufficient to assume that {\em space-time arguments of the operators which a time-normal product comprises are confined to the union of the past light cones of its arguments.\/}
Without assumptions about quantum dynamics, eq.\ (\ref{eq:39DA}) explicitly violates this condition. So, in Fig.\ \ref{fig:RelCaus}, the integration path extends into the future beyond the point ${\mbox{\rm\boldmath$r$}}_2,t_2$. The minimal dynamical assumption one has to make is that two operators commute if their arguments are separated by a space-like interval (it is one of Wightman's axioms \cite{Wightman}). Such commutativity holds for ${\hat{\mathcal A}}_{1}({\mbox{\rm\boldmath$r$}}_1,t_1)$ and ${\hat{\mathcal A}}_{2}({\mbox{\rm\boldmath$r$}}_2,t_2')$ with $\tilde t<t'_2<t_1$, where $\tilde t$ is the time when the integration path exits the past light cone of ${\mbox{\rm\boldmath$r$}}_1,t_1$ (Fig.\ \ref{fig:RelCaus}). With this observation, 
the contribution from the dashed section of the integration path in Fig.\ \ref{fig:RelCaus} reduces to, (cf.\ eq.\ (\ref{eq:24CK}))
{\begin{multline}\hspace{0.4\columnwidth}\hspace{-0.4\twocolumnwidth} 
\int_{\tilde t}^{t_1} dt'_2
\ensuremath{\big[
\delta ^{(+)}(t_2-t'_2) 
+ 
\delta ^{(-)}(t_2-t'_2)
\big]} {\hat{\mathcal A}}_{2}({\mbox{\rm\boldmath$r$}}_2,t'_2){\hat{\mathcal A}}_{1}({\mbox{\rm\boldmath$r$}}_1,t_1) 
\\ = \theta(t_2-\tilde t){\hat{\mathcal A}}_{2}({\mbox{\rm\boldmath$r$}}_2,t_2){\hat{\mathcal A}}_{1}({\mbox{\rm\boldmath$r$}}_1,t_1) 
. 
\hspace{0.4\columnwidth}\hspace{-0.4\twocolumnwidth} 
\label{eq:40DB} 
\end{multline}}%
The offending contribution from $t_2'>t_2$ cancelled. Similar arguments apply to the second term in (\ref{eq:39DA}) \footnote{The contribution to (\ref{eq:37CY}) thus comes from the points ${\mbox{\rm\boldmath$r$}}_1,t_1$, ${\mbox{\rm\boldmath$r$}}_2,t_2$ themselves plus the {\em intersection\/} of their past light cones. This is a stronger result than the relativistic ``no-peep-into-the-future'' theorem, but it may well be an artifact of the product of two operators.}. 

{\em Conclusion.---\/} We have demonstrated, both by example and by a formal proof, that causality violations, occuring if Kelley-Kleiner's definition of the time-normal operator ordering is taken outside the rotating wave approximation, are eliminated by putting the amended definition of Refs.\ \cite{APII,APIII} to use. Relativistic causality was verified for a time-normal average of two operators, while extention to more than two operators and, much more importantly, to renormalised theories remains subject to further work. 

{\em Acknowledgements.---\/} The authors thank A.Kaplan and W.P.Schleich for enlightening discussions, and D.Greenberger for comments on the manuscript. 
Support of SFB/TRR 21 and of the Humboldt Foundation is gratefully acknowledged. 
 
\end{document}